\documentclass[oneside,magyar,english]{amsart}
\usepackage[T1]{fontenc}
\usepackage[latin2,latin9]{inputenc}
\usepackage{color}
\usepackage{babel}
\usepackage{units}
\usepackage{mathrsfs}
\usepackage{amsthm}
\usepackage{amsbsy}
\usepackage{amstext}
\usepackage{amssymb}
\usepackage{xargs}[2008/03/08]
\usepackage[unicode=true,pdfusetitle,
 bookmarks=true,bookmarksnumbered=false,bookmarksopen=false,
 breaklinks=false,pdfborder={0 0 1},backref=false,colorlinks=true]
 {hyperref}
\hypersetup{
 citecolor=blue}

\makeatletter
\numberwithin{equation}{section}
\numberwithin{figure}{section}

\@ifundefined{date}{}{\date{}}
\usepackage{mathrsfs}
\usepackage{upgreek}

\makeatother

\begin{document}
\global\long\def\set#1#2{\left\{  #1\, |\, #2\right\}  }

\selectlanguage{magyar}%
\inputencoding{latin2}\global\long\def\smat#1#2#3#4{\bigl(\begin{smallmatrix}#1\,#2\cr\cr#3\,#4\end{smallmatrix}\bigr)}

\selectlanguage{english}%
\inputencoding{latin9}\global\long\def\inv{^{\textrm{\,-}{\scriptscriptstyle 1}}}

\global\long\def\map#1#2#3{#1\!:\!#2\!\rightarrow\!#3}

\global\long\def\Map#1#2#3#4#5{\begin{split}#1:#2  &  \rightarrow#3\\
#4  &  \mapsto#5 
\end{split}
 }

\global\long\def\cyc#1{\mathbb{Q}\left[\zeta_{#1}\right]}

\global\long\def\ZN#1{\left(\mathbb{Z}/#1\mathbb{Z}\right)^{\times}}

\global\long\def\Mod#1#2#3{#1\equiv#2\, \left(\mathrm{mod}\, \, #3\right)}

\global\long\def\aut#1{\mathrm{Aut}\!\left(#1\right)}

\global\long\def\End#1{\mathrm{End}\!\left(#1\right)}

\global\long\def\gl#1#2{\mathrm{GL}_{#1}\left(#2\right)}

\global\long\def\tr#1#2{\textrm{Tr}_{_{#1}}\left(#2\right)}

\global\long\def\sn#1{\mathbb{S}_{#1}}

\global\long\def\msp{\mathscr{V}}

\global\long\def\csp{\mathscr{W}}

\global\long\def\mb{\mathfrak{B}}

\global\long\def\mbv{\mathfrak{b}}

\global\long\def\FA{\mathbb{Z}\oplus\mathbb{Z}}

\global\long\def\FB{\mathbf{H}}

\global\long\def\FC{\mathrm{SL}_{2}\!\left(\mathbb{Z}\right)}

\global\long\def\FD{\ell_{i}}

\global\long\def\cw#1{\mathtt{h}_{#1}}

\global\long\def\ex#1{\mathtt{e}^{2\mathtt{i}\pi#1}}

\newcommandx\exi[3][usedefault, addprefix=\global, 1=, 2=]{\mathtt{e}^{\mathtt{{\scriptscriptstyle \textrm{#1}}}\frac{#2\pi\mathtt{i}}{#3}}}

\global\long\def\tw#1#2#3{\mathfrak{#1}_{#3}^{{\scriptscriptstyle \left(\!#2\!\right)}}}

\global\long\def\FE#1{\mathscr{L}\!\left(#1\right)}

\global\long\def\FF#1{\mathscr{C}\!\left(#1,R\right)}

\global\long\def\FG#1{#1^{*}}

\global\long\def\FH{}

\global\long\def\FI#1{#1_{{\scriptscriptstyle \pm}}}

\global\long\def\FJ#1#2#3{\uptheta\!\Bigl[\!{#1\atop #2}\!\Bigr]\!\left(#3\right)}

\title{Character relations and replication identities in 2d Conformal Field
Theory}

\author{P. Bantay}

\curraddr{Institute for Theoretical Physics, Eötvös Loránd University, H-1117
Budapest, Pázmány P. s. 1/A}

\email{bantay@poe.elte.hu}


\thanks{Work supported by grant OTKA 79005.}
\begin{abstract}
We study replication identities satisfied by conformal characters
of a 2D CFT, providing a natural framework for a physics interpretation
of the famous Hauptmodul property of Monstrous Moonshine, and illustrate
the underlying ideas in simple cases.
\end{abstract}

\maketitle

\section{Introduction}

The remarkable interaction between mathematics and physics around
the turn of the century has been to a large part spurred by the new
mathematical structures underlying String Theory \cite{GSW,Polch},
leading to such interesting new mathematical concepts as Vertex Operator
Algebras \cite{Borcherds1,FLM1} and Modular Tensor Categories \cite{Turaev,Bakalov-Kirillov}.
These developments in turn were strongly influenced by Monstrous Moonshine
\cite{Thompson1,Thompson2,Convay-Norton}, the amazing connection
between the representation theory of the Monster $\mathbb{M}$, the
largest sporadic finite simple group, with the classical theory of
modular forms. Actually, VOA theory grew out from the need to provide
a conceptual explanation of Moonshine. 

It has been recognized pretty early \cite{Dixon-Ginsparg-Harvey}
that, to a large extent, the Moonshine conjectures find a natural
physics explanation by interpreting the relevant quantities as describing
string propagation in a suitable (rather exotic) background, the Moonshine
orbifold, obtained as the result of orbifolding \cite{Dixon_orbifolds1}
the Moonshine module by the Monster. From this point of view, many
strange-looking properties \cite{generalizedMoonshine} of the Thompson-McKay
series involved in Moonshine follow from general physical principles,
with one notable exception: the so-called Hauptmodul property, which
states (roughly speaking) that Thompson-McKay series generate the
field of meromorphic functions of suitable genus zero Riemann surfaces,
does not find any obvious interpretation from a physics perspective
\cite{25years,Gannon2010}.

There has been several attempts to remedy this situation and find
a physics explanation of the Hauptmodul property, see e.g. \cite{Duncan2010,Tuite1995,Tuite2010},
but none proposed to this date seems completely satisfactory. The
aim of the present paper is to present a new approach to the problem,
based on the notion of character relations and replication identities,
which generalizes to arbitrary 2D Conformal Field Theories \cite{BPZ,DiFrancesco-Mathieu-Senechal},
and which provides an equivalent formulation of the Hauptmodul property
in the special case of the Moonshine orbifold. Roughly speaking, this
approach relates the Hauptmodul property to symmetries of second quantized
string propagation \cite{elliptic_genera} on the Moonshine orbifold.
While the precise nature of these symmetries is still unclear (because
identifying them would require a thorough analysis of the higher symmetric
products of the Moonshine orbifold, a pretty challenging task in view
of the intricate computations involved), the above identification
could prove to be a first step in a better understanding of the problem.
That the above approach can be made to work is demonstrated in the
comparatively much simpler case of the Ising model, where the analysis
can be explicitly performed (at least for low degrees), and the resulting
replication identities related precisely to actual symmetries of symmetric
products.

\section{Conformal characters, the modular representation and character relations}

Among the important characteristics of a 2D CFT \cite{BPZ,DiFrancesco-Mathieu-Senechal},
a prominent role is played by the conformal characters of the 'primaries',
the trace functions of irreducible modules in the language of ($C_{2}$-cofinite
rational) Vertex Operator Algebras. As a consequence of conformal
symmetry, the chiral symmetry algebra contains the Virasoro algebra,
whose zero mode $L_{0}$ plays the role of (chiral) Hamiltonian. The
commutation rules of the Virasoro generators imply that, in each irreducible
module separately, the eigenvalues of $L_{0}$ are integrally spaced,
hence the spectrum of $L_{0}$ can be characterized by specifying
the lowest eigenvalue, called the conformal weight of the primary,
and the generating function of the eigenvalue multiplicities. For
a primary $p$ of conformal weight $\cw p$, the conformal character
reads
\begin{equation}
\chi_{p}\!\left(q\right)=q^{{\scriptscriptstyle \nicefrac{-c}{24}}}\sum_{n=0}^{\infty}d_{n}q^{n+\cw p}\label{eq:chardef}
\end{equation}
where $d_{n}$ denotes the multiplicity of $n\!+\!\cw p$ as an eigenvalue
of $L_{0}$ and $c$ the central charge of the model. One can show
that the above (fractional) power series is absolutely convergent
in the disk $\bigl|q\bigr|<1$, hence defines an analytic function
there.

Besides characterizing the spectrum of $L_{0}$ in the irreducible
modules, the conformal characters also provide the basic building
blocks of the torus partition function. In the simplest case of diagonal
theories, the torus partition function reads
\begin{equation}
Z\!\left(\tau,\overline{\tau}\right)=\sum_{p}\bigl|\chi_{p}\!\left(\ex{\tau}\right)\bigr|^{2}\label{eq:partfun}
\end{equation}
where $\tau$ denotes the modular parameter of the torus, and the
sum runs over all primaries; more generally, the torus partition function
is a sesquilinear combination of the conformal characters. Combining
this observation with the invariance \cite{modinv} of the torus partition
function under modular transformations i.e. transformations of the
modular parameter $\tau$ that do not change the conformal equivalence
class, one arrives at the conclusion that the modular group $\FC$
is represented on the linear span of the characters, i.e. for any
$\smat abcd\!\in\!\FC$ there exists a unitary representation matrix
$M\!=\!\rho\smat abcd$ such that 
\begin{equation}
\chi_{p}\!\left(\frac{a\tau\!+\!b}{c\tau\!+\!d}\right)=\sum_{s}M{}_{ps}\chi_{s}\!\left(\tau\right)\label{eq:transrule}
\end{equation}

Two remarks are in order here: first, the modular representation is
actually a \emph{matrix} representation, meaning that each individual
modular matrix element has an invariant meaning. This is particularly
clear when considering Verlinde's celebrated formula \cite{Verlinde1988}
expressing the fusion rules of the theory in terms of modular matrix
elements, or its various generalizations \cite{Moore-Seiberg,Bantay2003c}.
From a technical point of view, this means that the linear space $\msp$
affording the modular representation comes equipped with a distinguished
basis $\mb\!=\!\left\{ \mbv_{p}\right\} $ labeled by the primaries,
and a different choice of basis would correspond to a different theory.

The second observation is that the transformation rule Eq.(\ref{eq:transrule})
does not always determine the modular representation matrices. The
reason for this is that the conformal characters, as functions of
the modular parameter $\tau$, are not necessarily linearly independent,
i.e. there may exist nontrivial relations of the form
\begin{equation}
\sum_{p}R_{p}\chi_{p}\!\left(\tau\right)=0\label{eq:charrel}
\end{equation}
with coefficients $R_{p}$ independent of $\tau$. The existence of
such nontrivial character relations is actually pretty common, e.g.
the characters of charge conjugate primaries are automatically equal
\begin{equation}
\chi_{\overline{p}}\!\left(\tau\right)=\chi_{p}\!\left(\tau\right)\label{eq:ccrel}
\end{equation}
As a consequence of the character relations, the linear span $\csp$
of the characters is usually only a subspace of $\msp$, and the individual
modular matrix elements cannot be determined from Eq.(\ref{eq:transrule}),
only suitable linear combinations of them. Actually, the example of
charge conjugation is a good indication for the origin of such character
relations: they are the reflections of (possibly hidden) global symmetries
of the theory\footnote{Indeed, character relations, as linear relations between suitable
(chiral) correlators, may be considered as Ward identities related
to some global symmetry. Of course, the precise nature of the relevant
symmetry might be pretty hard to pin down.}.

To illustrate this last point, let us consider the orbifold line of
$c\!=\!1$ theories \cite{Ginsparg1988}. It is well known that, at
compactification radii for which $N\!=\!2r_{\mathtt{orb}}^{2}$ is
an integer, these theories have exactly $N\!+\!7$ primary fields
with conformal characters
\begin{align}
\FI u\!\left(\tau\right) & =\frac{1}{2\eta\!\left(\tau\right)}\theta_{3}\!\left(2N\tau\right)\pm\sqrt{\frac{\eta}{2\theta_{2}}}\!\left(\tau\right)=\frac{1}{2\eta\!\left(\tau\right)}\left\{ \FJ 00{2N\tau}\pm\theta_{4}\!\left(2\tau\right)\right\} \nonumber \\
\chi_{k}\!\left(\tau\right) & =\frac{1}{\eta\!\left(\tau\right)}\FJ{\frac{k}{2N}}0{2N\tau}\textrm{ ~for }k\!=\!1,\ldots,N-1\nonumber \\
\FI{\phi}\!\left(\tau\right) & =\frac{1}{2\eta\!\left(\tau\right)}\theta_{2}\!\left(2N\tau\right)=\frac{1}{2\eta\!\left(\tau\right)}\FJ{\frac{1}{2}}0{2N\tau}\label{eq:atrels}\\
\FI{\sigma}\!\left(\tau\right) & =\frac{1}{2}\left\{ \sqrt{\frac{\eta}{\theta_{4}}}\!\left(\tau\right)+\sqrt{\frac{\eta}{\theta_{3}}}\!\left(\tau\right)\right\} =\frac{1}{2\eta\!\left(\tau\right)}\left\{ \theta_{2}\!\left(\frac{\tau}{2}\right)+\exi[-]{24}\theta_{2}\!\left(\!\frac{\tau\!+\!1}{2}\!\right)\right\} \nonumber \\
\FI{\tau}\!\left(\tau\right) & =\frac{1}{2}\left\{ \sqrt{\frac{\eta}{\theta_{4}}}\!\left(\tau\right)-\sqrt{\frac{\eta}{\theta_{3}}}\!\left(\tau\right)\right\} =\frac{1}{2\eta\!\left(\tau\right)}\left\{ \theta_{2}\!\left(\frac{\tau}{2}\right)-\exi[-]{24}\theta_{2}\!\left(\!\frac{\tau\!+\!1}{2}\!\right)\right\} \nonumber 
\end{align}
where
\begin{equation}
\FJ ab{\tau}=\sum_{n\in\mathbb{Z}}\mathtt{e}^{\mathtt{i}\pi\tau\left(n-a\right)^{2}}\mathtt{e}^{-2\pi\mathtt{i}bn}\label{eq:thetadef}
\end{equation}
and
\begin{equation}
\eta\!\left(\tau\right)=q^{\frac{1}{24}}\prod_{n=1}^{\infty}\left(1-q^{n}\right)\label{eq:etadef}
\end{equation}
denotes Dedekind's eta function (with $q\!=\!\ex{\tau}$), while
\begin{alignat*}{2}
\theta_{2} & =\FJ{\frac{1}{2}}0{\tau}~ & =2q^{\nicefrac{1}{8}}\prod_{n=1}^{\infty}\left(1-q^{n}\right)\left(1+q^{n}\right)^{2}\\
\theta_{3} & =\FJ 00{\tau}~ & =~\prod_{n=1}^{\infty}\left(1-q^{n}\right)\left(1+q^{n-\nicefrac{1}{2}}\right)^{2}\\
\theta_{4} & =\FJ 0{\frac{1}{2}}{\tau}~ & =~\prod_{n=1}^{\infty}\left(1-q^{n}\right)\left(1-q^{n-\nicefrac{1}{2}}\right)^{2}
\end{alignat*}
are the classical theta functions of Jacobi. 

Let's restrict our attention to the models with even $N$. Since charge
conjugation is trivial in this case, the obvious character relations
\begin{align}
\phi_{{\scriptscriptstyle -}} & =\phi_{{\scriptscriptstyle +}}\nonumber \\
\sigma_{{\scriptscriptstyle -}} & =\sigma_{{\scriptscriptstyle +}}\label{eq:ATcrel}\\
\tau_{{\scriptscriptstyle -}} & =\tau_{{\scriptscriptstyle +}}\nonumber 
\end{align}
must have a different origin: they are a manifestation of the dihedral
$\mathbb{D}_{4}$ symmetry underlying these models \cite{Ginsparg1988},
which follows from the fact that the orbifold line may be obtained
as the conformal limit of Ashkin-Teller models, i.e. two Ising spins
coupled locally via their energy density. Clearly, the transformations
that flip each Ising spin separately, together with the one that exchanges
the two, form a $\mathbb{D}_{4}$ symmetry group, explaining the above
character relations. In case $N\!=\!4$ (corresponding to the 4-state
Potts model) this symmetry is extended to a full $\mathbb{S}_{4}$,
resulting in the extra character relations
\begin{align}
\phi_{{\scriptscriptstyle \pm}} & =u_{{\scriptscriptstyle -}}\nonumber \\
\chi_{1} & =\sigma_{{\scriptscriptstyle \pm}}\label{eq:at4crel}\\
\chi_{3} & =\tau_{{\scriptscriptstyle \pm}}\nonumber 
\end{align}

Another interesting case is that of $N\!=\!16$, when the generic
character relations Eq.(\ref{eq:ATcrel}) get supplemented by
\begin{align}
\chi_{8}-\phi_{{\scriptscriptstyle \pm}} & =u_{{\scriptscriptstyle -}}\nonumber \\
\chi_{2}+\chi_{14} & =\sigma_{{\scriptscriptstyle \pm}}\label{eq:at16crel}\\
\chi_{6}+\chi_{10} & =\tau_{{\scriptscriptstyle \pm}}\nonumber 
\end{align}
More generally, such extra character relations occur whenever $N$
is the square of an even integer, $N\!=\!\left(2n\right)^{2}$, when
one has
\begin{align}
\sum_{k=1}^{n-1}(-1)^{k-1}\chi_{4nk}-(-1)^{n}\phi_{{\scriptscriptstyle \pm}} & =u_{{\scriptscriptstyle -}}\nonumber \\
\sum_{k=0}^{\left[\frac{n-1}{2}\right]}\left\{ \chi_{n(8k+1)}+\chi_{n(8k+7)}\right\}  & =\sigma_{{\scriptscriptstyle \pm}}\label{eq:atcrel}\\
\sum_{k=0}^{\left[\frac{n-1}{2}\right]}\left\{ \chi_{n(8k+3)}+\chi_{n(8k+5)}\right\}  & =\tau_{{\scriptscriptstyle \pm}}\nonumber 
\end{align}
as a consequence of the general identity
\begin{equation}
\sum_{k=0}^{N-1}\ex{\frac{kb}{N}}\FJ{a+\frac{k}{N}}0{\tau}=\FJ{-Na}{\frac{b}{N}}{\frac{\tau}{N^{2}}}\label{eq:thetaid}
\end{equation}
valid for integer $b$ and $N$, as well as the theta relations\footnote{An interesting consequence of Eq.(\ref{eq:atcrel}) is that in this
case the characters of the orbifold can be expressed as linear combinations
of the characters of the original theory, i.e. the compactified boson
at radius $r\!=\!\sqrt{2}n$.}
\begin{align}
\theta_{4}\!\left(2\tau\right) & =\sqrt{\theta_{3}\!\left(\tau\right)\theta_{4}\!\left(\tau\right)}\nonumber \\
\theta_{2}\!\left(\frac{\tau}{2}\right) & =\sqrt{2\theta_{2}\!\left(\tau\right)\theta_{3}\!\left(\tau\right)}\label{eq:thetarel}\\
\theta_{2}\!\left(\!\frac{\tau\!+\!1}{2}\!\right) & =\mathtt{e}^{\frac{\mathtt{i}\pi}{16}}\sqrt{2\theta_{2}\!\left(\tau\right)\theta_{4}\!\left(\tau\right)}\nonumber 
\end{align}
The origin of these extra relations Eq.(\ref{eq:atcrel}) may be traced
back to the fact that the corresponding models may be constructed
as dihedral orbifolds of the compactified boson at radius $r\!=\!\nicefrac{1}{\sqrt{2}}$
\cite{Ginsparg1988}. 

From a technical point of view, nontrivial character relations indicate
that the modular representation $\rho$ is reducible. Indeed, as a
consequence of the $\tau$ independence of the coefficients $R_{p}$
in Eq.(\ref{eq:charrel}), the linear span $\csp$ of the characters
(considered as a subspace of $\msp$) is invariant under $\rho$.
In particular, this means that in order to fully characterize the
modular properties of the characters, it is not enough to specify
the matrix representation $\rho$, but one should amend this by a
description of the invariant subspace $\csp$ (e.g. by specifying
a basis of it). Formally, one could think that this last step can
be avoided by directly reducing the modular representation to the
invariant subspace $\csp$: after all, this subspace is the linear
span of the conformal characters, thus it contains all the physically
relevant information. But this is far from being true. For example,
application of Verlinde's formula \cite{Verlinde1988,Moore-Seiberg},
one of the cornerstones of the whole theory, necessitates the consideration
of the full modular representation, with all individual matrix elements.
Similarly, computation of Frobenius-Schur indicators \cite{Bantay1997a},
or the application of the trace identities of \cite{Bantay2003c}
require the knowledge of each matrix element separately.

\section{Symmetric products and replication identities}

Consider a system made up of $n$ identical subsystems, each described
by the same CFT $\mathcal{C}$. The whole system will be still conformally
invariant, described by the $n$-fold tensor power of $\mathcal{C}$,
and any permutation of the identical subsystems will leave the whole
system invariant. Consequently, for any permutation group $\Omega\!<\!\sn n$
of degree $n$, one could consider the permutation orbifold\footnote{The origin of the wreath product notation for permutation orbifolds
is explained in \cite{Bantay1998a}.} $\mathcal{C}\wr\Omega$ obtained by orbifolding the tensor power
by the twist group $\Omega$ \cite{Klemm-Schmidt,Borisov-Halpern-Schweigert,Bantay1998a}.
Because of the universal nature of the action of $\Omega$, all relevant
quantities (like correlation and partition functions, fusion rules,
modular matrix elements, etc.) of $\mathcal{C}\!\wr\!\Omega$ may
be expressed in terms of the relevant quantities of $\mathcal{C}$,
namely as polynomial expressions of these quantities evaluated on
suitable $n$-sheeted covering surfaces of the world sheet, see \cite{Bantay2001,Bantay2002}
for details. In particular, the conformal characters of the permutation
orbifold are completely determined by those of $\mathcal{C}$ and
the twist group $\Omega$ \cite{Bantay1998a}. We note that all relevant
relations can be subsumed under a general group theoretic construct,
the orbifold transform, described in detail in \cite{Bantay2008a}.

A particularly interesting case is when the twist group $\Omega$
is maximal, i.e. when $\Omega$ is the full symmetric group $\sn n$
of degree $n$: the resulting permutation orbifold $\mathcal{C}\!\wr\!\sn n$
is called the $n$-th symmetric product of $\mathcal{C}$, and plays
an important role in the description of second quantized strings \cite{elliptic_genera,Dijkgraaf_disctors,Bantay2003a}.
The analysis of symmetric products is greatly simplified by the exponential
identity \cite{Bantay2008a}, a general combinatorial identity satisfied
by the orbifold transform, which provides closed expressions for the
characteristic quantities of symmetric products. 

According to the general theory \cite{Bantay1998a}, the conformal
characters (evaluated at some specific modulus $\tau$) of the $n$-fold
symmetric product $\mathcal{C}\wr\sn n$ may be expressed as polynomial
expressions of the conformal characters of $\mathcal{C}$ evaluated
on the different $n$-sheeted (unbranched) coverings of a torus with
modulus $\tau$. But all theses coverings have genus $1$, hence each
connected component is itself a torus of modulus
\begin{equation}
\frac{a\tau+b}{d}\label{eq:covertau}
\end{equation}
for suitable non-negative integers $a,b,d$ characterizing the relevant
covering. The precise form of the polynomial expressions is irrelevant
at this point, the only thing to note is that all possible coverings
occur in the process. This means that a character relation of the
symmetric product $\mathcal{C}\wr\sn n$ is nothing but a polynomial
relation between quantities of the form
\[
\chi_{p}\!\left(\!\frac{a\tau+b}{d}\!\right)
\]
We shall call such relations replication identities, because in the
specific case of the Moonshine orbifold they yield precisely the replication
formulas satisfied by the generalized Thompson-McKay series. It should
be emphasized that the above notion of replication identities is pretty
general, far from being confined to derivatives of the Moonshine module
or to rational conformal models.

As explained in the previous section, character relations for a given
CFT are usually reflections of outer symmetries relating different
irreducible modules of the chiral algebra. Consequently, one may view
replication identities as an indication to the existence of suitable
symmetries of the higher symmetric products. To illustrate the above
ideas, let us consider the Ising model, i.e. the Virasoro minimal
model of central charge $c\!=\!\nicefrac{1}{2}$. In this case there
are three primary fields, $\boldsymbol{0},~\boldsymbol{\epsilon}$
and $\boldsymbol{\sigma}$, of respective conformal weights $0,\nicefrac{1}{2}$
and $\nicefrac{1}{16}$, with conformal characters
\begin{align}
\chi_{\boldsymbol{{\scriptscriptstyle 0}}} & =\frac{1}{2}\!\left(\sqrt{\frac{\theta_{3}}{\eta}}+\sqrt{\frac{\theta_{4}}{\eta}}~\right)\nonumber \\
\chi_{\boldsymbol{\epsilon}} & =\frac{1}{2}\!\left(\sqrt{\frac{\theta_{3}}{\eta}}-\sqrt{\frac{\theta_{4}}{\eta}}~\right)\label{eq:isingchars}\\
\chi_{\boldsymbol{\sigma}} & =\sqrt{\frac{\theta_{2}}{2\eta}}\nonumber 
\end{align}
 Note that
\begin{alignat}{2}
\sqrt{\frac{\theta_{3}}{\eta}}= & ~{\displaystyle q^{\textrm{-}1/48}\prod_{n=0}^{\infty}}\left(1+q^{n+\frac{1}{2}}\right) & ~=\frac{\eta\!\left(\tau\right)^{2}}{\eta\!\left(\frac{\tau}{2}\right)\eta\!\left(2\tau\right)}\nonumber \\
\sqrt{\frac{\theta_{4}}{\eta}}= & ~q^{\textrm{-}1/48}\prod_{n=0}^{\infty}\left(1-q^{n+\frac{1}{2}}\right) & =\frac{\eta\!\left(\frac{\tau}{2}\right)}{\eta\!\left(\tau\right)}~~~~~~~~\label{eq:thetaprod}\\
\sqrt{\frac{\theta_{2}}{2\eta}}= & ~{\displaystyle q^{1/24}~\prod_{n=1}^{\infty}}\left(1+q^{n}\right) & =\frac{\eta\!\left(2\tau\right)}{\eta\!\left(\tau\right)}~~~~~~~~\nonumber 
\end{alignat}

The modular representation, characterized by the matrix
\begin{equation}
S=\frac{1}{2}\left(\begin{array}{rrr}
1 & 1 & \sqrt{2}\\
1 & 1 & -\sqrt{2}\\
\sqrt{2} & -\sqrt{2} & 0
\end{array}\right)\label{eq:isingS}
\end{equation}
is irreducible, hence has no non-trivial invariant subspace; consequently,
the conformal characters of Ising are linearly independent. 

As a consequence of the identities
\begin{align}
\sqrt{\frac{\theta_{4}\!\left(2\tau\right)}{\eta\!\left(2\tau\right)}} & =\frac{\sqrt{\theta_{3}\!\left(\tau\right)\theta_{4}\!\left(\tau\right)}}{\eta\!\left(\tau\right)}\nonumber \\
\sqrt{\frac{\theta_{2}\!\left(\frac{\tau}{2}\right)}{\eta\!\left(\frac{\tau}{2}\right)}} & =\frac{\sqrt{\theta_{2}\!\left(\tau\right)\theta_{3}\!\left(\tau\right)}}{\eta\!\left(\tau\right)}\label{eq:thetarel2}\\
\exi[-]{24}\sqrt{\frac{\theta_{2}\!\left(\frac{\tau+1}{2}\right)}{\eta\!\left(\frac{\tau+1}{2}\right)}} & =\frac{\sqrt{\theta_{2}\!\left(\tau\right)\theta_{4}\!\left(\tau\right)}}{\eta\!\left(\tau\right)}\nonumber 
\end{align}
that follow easily from Eqs.(\ref{eq:thetarel}), one gets that
\begin{align}
\chi_{\boldsymbol{{\scriptscriptstyle 0}}}\!\left(2\tau\right)-\chi_{\boldsymbol{\epsilon}}\!\left(2\tau\right) & =\chi_{\boldsymbol{{\scriptscriptstyle 0}}}\!\left(\tau\right)^{2}-\chi_{\boldsymbol{\epsilon}}\!\left(\tau\right)^{2}\nonumber \\
\chi_{\boldsymbol{\sigma}}\!\left(\frac{\tau}{2}\right) & =\frac{\chi_{\boldsymbol{{\scriptscriptstyle 0}}}\!\left(\tau\right)\!+\!\chi_{\boldsymbol{\epsilon}}\!\left(\tau\right)}{2}\chi_{\boldsymbol{\sigma}}\!\left(\tau\right)\label{eq:Is2rels}\\
\chi_{\boldsymbol{\sigma}}\!\left(\!\frac{\tau\!+\!1}{2}\!\right) & =\exi{24}\frac{\chi_{\boldsymbol{{\scriptscriptstyle 0}}}\!\left(\tau\right)\!-\!\chi_{\boldsymbol{\epsilon}}\!\left(\tau\right)}{2}\chi_{\boldsymbol{\sigma}}\!\left(\tau\right)\nonumber 
\end{align}
These are prime examples of replication identities, involving values
of characters on different covering surfaces. Consequently, they should
be related to character relations, and ultimately to symmetries of
symmetric products of the Ising model. Let's see how this comes about!

\global\long\def\chipq#1#2{\boldsymbol{\upphi}_{{\scriptscriptstyle \left\langle #1,#2\right\rangle }}}

According to the general theory \cite{Bantay1998a}, the 2-fold symmetric
product of Ising has central charge $c\!=\!1$ (twice the central
charge of the Ising model) and a total of $\frac{3\left(3+7\right)}{2}\!=\!15$
primary fields, whose conformal characters read
\begin{align}
\chipq pq\!\left(\tau\right) & =\chi_{p}\!\left(\tau\right)\chi_{q}\!\left(\tau\right)~~~\enspace~\quad~\quad~~~\textrm{for }~p\neq q\nonumber \\
\tw u{\pm}p\!\left(\tau\right) & =\frac{1}{2}\left\{ \chi_{p}\!\left(\tau\right)^{2}\pm\chi_{p}\!\left(2\tau\right)\!\right\} \label{eq:Is2chars}\\
\tw t{\pm}p\!\left(\tau\right) & =\frac{1}{2}\left\{ \chi_{p}\!\left(\frac{\tau}{2}\right)\pm\mathtt{e}^{-\mathtt{i}\pi\left(\cw p-\nicefrac{1}{48}\right)}\chi_{p}\!\left(\!\frac{\tau\!+\!1}{2}\!\right)\!\right\} \nonumber 
\end{align}
for $p,q\!\in\!\left\{ \boldsymbol{0},\boldsymbol{\epsilon},\boldsymbol{\sigma}\right\} $,
with $\cw p$ denoting the conformal weight of the primary $p$. By
inspecting the $q$-expansions of these characters, one arrives at
the character relations
\begin{align}
\tw u-{\boldsymbol{{\scriptscriptstyle 0}}}\!\left(\tau\right) & =\tw u-{\boldsymbol{\epsilon}}\!\left(\tau\right)\nonumber \\
\chipq{\boldsymbol{0}}{\boldsymbol{\sigma}}\!\left(\tau\right) & =\tw t+{\boldsymbol{{\scriptscriptstyle \sigma}}}\!\left(\tau\right)\label{eq:Is2charrels}\\
\chipq{\boldsymbol{\boldsymbol{\epsilon}}}{\boldsymbol{\sigma}}\!\left(\tau\right) & =\tw t-{\boldsymbol{{\scriptscriptstyle \sigma}}}\!\left(\tau\right)\nonumber 
\end{align}
which reduce to
\begin{align}
\chi_{\boldsymbol{{\scriptscriptstyle 0}}}\!\left(\tau\right)^{2}-\chi_{\boldsymbol{{\scriptscriptstyle 0}}}\!\left(2\tau\right) & =\chi_{\boldsymbol{\epsilon}}\!\left(\tau\right)^{2}-\chi_{\boldsymbol{\epsilon}}\!\left(2\tau\right)\nonumber \\
\chi_{\boldsymbol{\sigma}}\!\left(\frac{\tau}{2}\right)+\exi[-]{24}\chi_{\boldsymbol{\sigma}}\!\left(\!\frac{\tau\!+\!1}{2}\!\right) & =\chi_{\boldsymbol{{\scriptscriptstyle 0}}}\!\left(\tau\right)\chi_{\boldsymbol{\sigma}}\!\left(\tau\right)\label{eq:Is2relsbis}\\
\chi_{\boldsymbol{\sigma}}\!\left(\frac{\tau}{2}\right)-\exi[-]{24}\chi_{\boldsymbol{\sigma}}\!\left(\!\frac{\tau\!+\!1}{2}\!\right) & =\chi_{\boldsymbol{\epsilon}}\!\left(\tau\right)\chi_{\boldsymbol{\sigma}}\!\left(\tau\right)\nonumber 
\end{align}
upon taking into account the expressions Eqs.(\ref{eq:Is2chars}).
Clearly, these are equivalent to the replication identities Eqs.(\ref{eq:Is2rels}),
and we see that, indeed, the latter are nothing but character relations
for the second symmetric product. What remains to do is to find out
which symmetries are responsible for this.

Since the full moduli space of $c\!=\!1$ conformal models is known,
it is a simple matter to identify the second symmetric product of
the Ising model: it lies on the orbifold line at radius $r_{\mathtt{orb}}\!=\!2$,
i.e. at $N\!=\!8$. Furthermore, it is an easy exercise to identify
the respective primary fields, in particular one gets\global\long\def\fid{\leftrightarrow}
\medskip{}
\begin{flalign}
u_{{\scriptscriptstyle +}} & \fid\tw u+{\boldsymbol{{\scriptscriptstyle 0}}} & \chi_{2} & \fid\tw u+{\boldsymbol{{\scriptscriptstyle \sigma}}} & \chi_{5} & \fid\tw t-{\boldsymbol{\epsilon}} & \phi_{{\scriptscriptstyle +}} & \fid\tw u-{\boldsymbol{{\scriptscriptstyle 0}}} & \phi_{{\scriptscriptstyle -}} & \fid\tw u-{\boldsymbol{\epsilon}}\nonumber \\
~~u_{{\scriptscriptstyle -}} & \fid\tw u+{\boldsymbol{\epsilon}} & \chi_{3} & \fid\tw t+{\boldsymbol{\epsilon}} & \chi_{6} & \fid\tw u-{\boldsymbol{{\scriptscriptstyle \sigma}}} & \sigma_{{\scriptscriptstyle +}} & \fid\chipq{\boldsymbol{0}}{\boldsymbol{\sigma}} & \sigma_{{\scriptscriptstyle -}} & \fid\tw t+{\boldsymbol{{\scriptscriptstyle \sigma}}}~~~~~\label{eq:at8fieldids}\\
\chi_{1} & \fid\tw t+{\boldsymbol{{\scriptscriptstyle 0}}} & \chi_{4} & \fid\chipq{\boldsymbol{0}}{\boldsymbol{\boldsymbol{\epsilon}}} & \chi_{7} & \fid\tw t-{\boldsymbol{{\scriptscriptstyle 0}}} & \tau_{{\scriptscriptstyle +}} & \fid\chipq{\boldsymbol{\boldsymbol{\epsilon}}}{\boldsymbol{\sigma}} & \tau_{{\scriptscriptstyle -}} & \fid\tw t-{\boldsymbol{{\scriptscriptstyle \sigma}}}\nonumber 
\end{flalign}

\medskip{}
But, as explained in the previous section, this model exhibits a dihedral
$\mathbb{D}_{4}$ symmetry as a consequence of its Ashkin-Teller origin,
leading to the character relations Eqs.(\ref{eq:ATcrel}), which,
taking into account the field identifications Eq.(\ref{eq:at8fieldids}),
yield precisely Eqs.(\ref{eq:Is2charrels}). In this case the analysis
of the relevant symmetries is relatively easy thanks to the identification
of the symmetric product as an Ashkin-Teller model, but the underlying
idea should be clear.

The third symmetric product of Ising has central charge $c\!=\!\nicefrac{3}{2}$,
and can be identified with an isolated $N\!=\!1$ superconformal model
\cite{Cappelli2001}, which has a total of $49$ primaries. This superconformal
model has $9$ independent character relations, but it turns out that
all of these follow from the replication identities (\ref{eq:Is2rels}).
New replication identities could come from the character relations
of the 4-fold symmetric product: unfortunately, this latter model
of central charge $c\!=\!2$ has $171$ different primary fields,
with $59$ independent relations between their characters, whose connection
to the symmetries of the model is far from being easy to determine.

\section{outlook and conclusion}

Trying to find a physics interpretation of the Hauptmodul property
of Monstrous Moonshine, we considered the question of character relations
and replication identities in Conformal Field Theory. Character relations
play an important role in understanding the structure of specific
models, and should be viewed as one of the basic ingredients (besides
the modular representation) to fully specify their modular properties,
while replication identities are nothing but character relations for
symmetric products. Since character relations can be traced back ultimately
to suitable symmetries of the model under study, replication identities
should correspond to symmetries of its symmetric products.

The Hauptmodul property of Monstrous Moonshine is a consequence of
the replication identities obeyed by the (generalized) Thompson-McKay
series. Based on this, we suggest that it is actually a manifestation
of the inherent symmetries of second quantized string propagation
on the Moonshine orbifold, the string background obtained by orbifolding
the Moonshine module by the Monster, and whose primary characters
are linear combinations of the Thompson-McKay series. Let us stress
that this approach does not give us an alternate proof of the Hauptmodul
property, just a possible physics interpretation for it. However,
if correct, it could have interesting consequences even from a purely
mathematical perspective, e.g. providing suitably generalized versions
of the replication identities for higher genus analogues of the Thompson-McKay
series.

While the arguments leading to the above could seem straightforward,
the actual implementation, i.e. the identification of the relevant
symmetries might be far from  simple. The proliferation of character
relations in higher symmetric products makes the analysis pretty difficult
even for the Ising model, and one should expect worse in more complicated
cases. But there are various arguments suggesting that, notwithstanding
all computational difficulties involved, the identification of the
relevant symmetries might be nevertheless carried out.

The first observation is that, for any two permutation groups $\Omega_{1}$
and $\Omega_{2}$ such that $\Omega_{1}$ is a subgroup of $\Omega_{2}$,
the character relations of the $\Omega_{1}$ permutation orbifold
are inherited by the $\Omega_{2}$ permutation orbifold. This is actually
the reason why it is sufficient to look only at symmetric products
when considering replication identities. Combining this with the obvious
embeddings of wreath products into symmetric groups and the transitivity
property of permutation orbifolds \cite{Bantay1998a,Bantay2002},
one can see that many of the replication identities of a given degree
are trivial consequences of lower degree ones, and in particular of
character relations, which are nothing but replication identities
of degree one. As a result, it is enough to understand the 'primitive'
identities that do not follow from identities coming from lower degrees,
and these are clearly much less abundant, hopefully forming a set
that can be dealt with. 

The second point is that one does not even need the precise identification
of all of the symmetries responsible for the primitive identities,
it is enough to identify only a generating set, which can turn out
to be pretty small. Since all replication identities for Moonshine
are known, this should simplify the job to a large extent. Of course,
even in case of a few generators the actual identification of the
relevant symmetries could require some ingenuity, but one could expect
that special properties of the Monster and the Moonshine module should
allow the use of ad hoc techniques to solve this problem: after all,
such considerations allow the determination of the character table
of the Monster (with cca. $10^{54}$ elements), while a brute force
computation for a group with only a few million of elements is already
a time and resource consuming task. 

Even if the above program can be completed and all relevant symmetries
responsible for the replication identities of the Moonshine orbifold
identified, there would still remain the question of what is so special
about this particular model. After all, while non-trivial replication
identities are not uncommon for rational models, they are usually
not restrictive enough to force the chiral characters to be actually
Hauptmoduls; this seems to be connected with a particularly high degree
of symmetry inherent to symmetric products of the Moonshine orbifold.
It would be interesting to find out other models that show similar
features, and whether this could be linked with other approaches \cite{Duncan2010,Tuite1995}
to the Hauptmodul property of Moonshine. We believe that further elaboration
of these issues could lead to a better understanding of the whole
subject.

\bibliographystyle{plain}
\phantomsection\addcontentsline{toc}{section}{\refname}\bibliography{C:/Users/mester/Documents/bibfiles/conformalcharacters,C:/Users/mester/Documents/bibfiles/modularfunctions,C:/Users/mester/Documents/bibfiles/CFT,C:/Users/mester/Documents/bibfiles/my,C:/Users/mester/Documents/bibfiles/math,C:/Users/mester/Documents/bibfiles/strings,C:/Users/mester/Documents/bibfiles/VOA,C:/Users/mester/Documents/bibfiles/orbifold_permutation,C:/Users/mester/Documents/bibfiles/phys,C:/Users/mester/Documents/bibfiles/Moonshine}

\end{document}